\documentclass[12pt,prd,tightenlines,nofootinbib,showpacs,showkeys]{revtex4}
\usepackage{bm}
\usepackage{graphics}
\usepackage{rotating}
\usepackage{epsfig}
\usepackage{amsmath}
\usepackage{amsfonts}
\usepackage{amssymb}
\usepackage{makeidx}

\begin{document}
\title{
Hyperfine structure of $S$-states in muonic ions of lithium, beryllium and boron.}
\author{A.~E.~Dorokhov\footnote{E-mail:~dorokhov@theor.jinr.ru}}
\affiliation{Joint Institute of Nuclear Research, BLTP, 141980, Moscow region, Dubna, Russia}
\author{\firstname{A.~A.} \surname{Krutov}}
\author{\firstname{A.~P.} \surname{Martynenko\footnote{E-mail:~a.p.martynenko@samsu.ru}}}
\author{\firstname{F.~A.} \surname{Martynenko}}
\author{\firstname{O.~S.} \surname{Sukhorukova}}
\affiliation{Samara University, 443086, Samara, Russia}

\begin{abstract}
We make precise calculation of hyperfine structure of $S$-states
in muonic ions of lithium, beryllium and boron in quantum 
electrodynamics. Corrections of orders $\alpha^5$ and $\alpha^6$
due to the vacuum polarization, nuclear structure and recoil in first and
second orders of perturbation theory are taken into account. 
We obtain estimates of the total values of hyperfine splittings
which can be used for a comparison with future experimental data. 
\end{abstract}

\pacs{31.30.jf, 12.20.Ds, 36.10.Ee}

\keywords{Hyperfine structure, muonic ions, quantum electrodynamics.}

\maketitle

\section{Introduction}

The hyperfine splitting (HFS) of 2S-state in muonic hydrogen was measured recently 
by the CREMA collaboration in \cite{crema1}:
\begin{equation}
\label{eq:1}
\Delta E^{hfs}_{exp}(2S)=22.8089(51)~meV.
\end{equation}
The theoretical value of the HFS of 2S-level, which was calculated with high accuracy as a result of taking into account numerous corrections for the vacuum polarization, structure and recoil of the nucleus, relativism agrees well with the value \eqref{eq:1}. At present, several experimental groups plan to measure the HFS of the ground state in muonic hydrogen with a record accuracy of 1 ppm. This will allow us to better study effects of the structure and polarizability of the proton. Since measurements have already been made of certain transition frequencies $(2P-2S)$ in muonic deuterium and muonic helium ion \cite{crema2,crema3,crema4}, it will apparently allow the experimental value of the HFS of 2S-state to be obtained in the near future for these muonic atoms.

The CREMA collaboration has obtained in recent years significantly new experimental results that helped to re-examine the problem of muon bound states, posed new questions to the theory that require additional investigation. One possible future activity of the CREMA collaboration may be connected with other muonic ions containing light nuclei of lithium, beryllium and boron. For these muonic ions, the description of the electromagnetic interaction of the few-nucleon systems is particularly important, and, consequently, the role of the effects of nuclear physics can be studied with greater accuracy. We may hope that the enormous interest, which experimental results of the CREMA collaboration have met over the past years can ultimately lead to a significant improvement in the theory of calculating the energy levels of muonic atoms.

In our previous work, we calculated the Lamb shift in the muonic ions of lithium, beryllium, and boron \cite{apm2016}. The purpose of this paper is to investigate the HFS of the S-states in these ions, that is, in the precise calculation of various corrections and obtaining reliable estimates for the HFS intervals, which could be used for comparison with experimental data. It should be noted that estimates of a number of important contributions to the HFS of ions have already been made in \cite{drake}. The initial parameters that determine the values of the corrections in the HFS of muonic ions, are the masses of the nuclei, their spins, magnetic moments and charge radii. Since we calculate the corrections immediately for
several nuclei, their parameters are presented in a separate Table~\ref{tb1} \cite{stone,marinova}.

A part of the Breit Hamiltonian, responsible for hyperfine splitting,
has a known form in the coordinate representation:
\begin{equation}
\label{eq:2}
\Delta V_B^{hfs}(r)=\frac{4\pi\alpha(1+a_\mu)\mu_N}{3m_1m_ps_2}({\bf s}_1{\bf s}_2)
\delta({\bf r}),
\end{equation}
where the masses of the muon and nuclear will be denoted further $m_1$, $m_2$,  
$m_p$ is the proton mass, $\mu_N$ is the nuclear magnetic moment in nuclear magnetons, 
$a_\mu$ is the muon anomalous magnetic moment (AMM), ${\bf s}_1$ and ${\bf s}_2$ are
the spins of a muon and nucleus.
The potential $\Delta V_B^{hfs}$ gives the main part of hyperfine splitting
of order $\alpha^4$ which is called the Fermi energy:
\begin{equation}
\label{eq:3}
E_F(nS)=\frac{2Z^3\alpha^4\mu^3\mu_N}{3m_1m_pn^3s_2}(2s_2+1),
\end{equation}
where $n$ is the principal quantum number, $\mu=m_1m_2/(m_1+m_2)$.
The factor $Z^3$ ($Z$ is the charge of the nucleus in units of the electron charge) 
in \eqref{eq:3} leads to essential increase of numerical values $E_F(nS)$
for the muonic ions of lithium, beryllium and boron in comparison with muonic hydrogen. 
The muon AMM $a_\mu$ is not included in \eqref{eq:3}.

\begin{table}[htbp]
\caption{The nucleus parameters of lithium, beryllium and boron.}
\label{tb1}
\bigskip
\begin{ruledtabular}
\begin{tabular}{|c|c|c|c|c|c|c|}   \hline
Nucleus & Spin &  Mass ,    & Magnetic dipole  & Charge radius, & Electroquadrupole  &  Magnetic octupole \\  
   &     & GeV     &   moment, nm   &   fm    & moment, fm$^2$   & moment, nm$\cdot$fm$^2$   \\    \hline
$^6_3Li$ & 1 & 5.60152 & 0.8220473(6) & $2.5890\pm 0.0390 $ & -0.083(8)  & 0 \\ \hline
$^7_3Li$ & 3/2 & 6.53383 & 3.256427(2) & $2.4440\pm 0.0420 $  & -4.06(8) &  7.5 \\ \hline
$^9_4Be$ & 3/2 & 8.39479 & -1.177432(3) & $2.5190\pm 0.0120 $ & 5.29(4)  &  4.1  \\ \hline
$^{10}_5B$ & 3 & 9.32699 & 0.8220473(6) & $2.4277\pm 0.0499 $ &  8.47(6) &  0  \\ \hline
$^{11}_5B$ & 3/2 & 10.25510 & 0.8220473(6) & $2.4060\pm 0.0294 $ &  4.07(3)& 7.8   \\ \hline
\end{tabular}
\end{ruledtabular}
\end{table}

The Fermi energy is obtained after averaging \eqref{eq:2} over the Coulomb wave functions.
In the case of $1S$- and $2S$-states they have the form:
\begin{equation}
\label{eq:4}
\psi_{100}(r)=\frac{W^{3/2}}{\sqrt{\pi}}e^{-Wr},~~~W=\mu Z\alpha,
\end{equation}
\begin{equation}
\label{eq:5}
\psi_{200}(r)=\frac{W^{3/2}}{2\sqrt{2\pi}}e^{-Wr/2}\left(1-\frac{Wr}{2}\right).
\end{equation}
The muon AMM correction to hyperfine splitting is presented separately in
Tables~\ref{tb2}-\ref{tb4} (line 2) taking experimental value of muon AMM \cite{MT}:
\begin{equation}
\label{eq:6}
\Delta E^{hfs}_{a_\mu}(nS)=a_\mu E_F(nS).
\end{equation}
Numerical value of relativistic correction of order $\alpha^6$ to HFS can be obtained by
means of known analytical expression from~\cite{egs,breit}:
\begin{equation}
\label{eq:7}
\Delta E^{hfs}_{rel}(nS)=\Biggl\{{{\frac{3}{2}(Z\alpha)^2E_F(1S)}\atop
{\frac{17}{8}(Z\alpha)^2E_F(2S)}}.
\end{equation}

Next, we investigate a number of basic corrections to the hyperfine structure
of $S$-states
in order to obtain acceptable total result. Numerical values of different corrections are
presented for definiteness with the accuracy $10^{-2}$ meV.

\begin{table}[htbp]
\caption{\label{tb2}Hyperfine splittings of S-states in muonic
ions $(\mu~^ 6_3Li)^{2+}$ and $(\mu~^ 7_3Li)^{2+}$.}
\bigskip
\begin{ruledtabular}
\begin{tabular}{|c|c|c|c|c|c|}  \hline
 No. & Contribution to the splitting &\multicolumn{2}{c|}{$(\mu~^ 6_3Li)^{2+}$,~meV}&
\multicolumn{2}{c|}{$(\mu~^ 7_3Li)^{2+}$, meV }   \\   \hline
  & &1S~~~~~~~ & 2S & 1S~~~~~~~& 2S \\   \hline
1 & Contribution of order $\alpha^4$, &  1416.07   &177.01 &5026.00 & 628.25  \\ 
  &  the Fermi energy    &     &      &       &    \\    \hline
2 & Muon AMM contribution  &  1.65   & 0.21& 5.87 &  0.73    \\  \hline
3 & Relativistic correction &  1.02     & 0.18 &3.62 &  0.64     \\
  & of order $\alpha^6$   &           &              &      &           \\   \hline  
4 & Nuclear structure correction & G;~-109.92 & G:~-13.74&G:~-369.25&G:~-46.16 \\ 
  &         of order $\alpha^5$        & U:~-112.02 &U:~-14.00 &U:~-376.31 & U:~-47.04  \\  \hline
5 & Nuclear structure and recoil & G:~-0.20 & G:~-0.03&G:~-30.67&G:~-3.83 \\    \hline  
6 &  Nuclear structure correction  & 3.35   &  0.34   & 10.67  &  1.08   \\
 & of order $\alpha^6$ in  $1\gamma$ interaction    &     &   &  &         \\   \hline
7 & Nuclear structure correction in second & -2.56  &-0.90  &-8.19 & -2.90         \\
 & order perturbation theory    &     &      &    &        \\   \hline
8 & Vacuum polarization contribution & 5.22& 0.67 &18.54  & 2.38      \\
 & of order $\alpha^5$ in first order PT &       &   &   &            \\   \hline
9 & Vacuum polarization contribution &12.05 &1.11  & 42.83 &  3.94     \\
 & of order $\alpha^5$ in second order PT &       &   &   &            \\   \hline
10 & Muon vacuum polarization contribution & 0.08& 0.01 &0.29  &  0.04     \\
 & of order $\alpha^6$ in first order PT &       &   &   &            \\   \hline
11 & Muon vacuum polarization contribution &0.09 &0.01  &0.31  & 0.04      \\
 & of order $\alpha^6$ in second order PT &       &   &   &            \\   \hline
12 & Vacuum polarization contribution   & 0.07 & 0.01 & 0.24 &  0.03    \\
 & of order $\alpha^6$ in first order PT  &      &      &   &      \\   \hline
13 & Vacuum polarization contribution  &0.14  &0.02  &0.53   &0.05     \\
 &  of order $\alpha^6$ in second order PT         &   &      &       &      \\   \hline
14 & Nuclear structure and vacuum & -1.62  & -0.20  & -5.85 & -0.73    \\
 &  polarization correction of order  $\alpha^6$ &     &    &   &    \\    \hline
15& Nuclear structure and muon vacuum  &  -0.14 & -0.02  &-0.51 & -0.06        \\
 & polarization correction of order  $\alpha^6$ &     &  &   &        \\    \hline
16 & Hadron vacuum polarization &  0.06   & 0.01  & 0.21 &  0.03     \\
 & contribution of order $\alpha^6$   &     &   &       &                  \\    \hline
17& Radiative nuclear finite size  & -0.34   &-0.04   &-1.24 &  -0.15     \\
 &  correction of order  $\alpha^6$    &     &   &   &          \\   \hline
&  Summary contribution  & 1325.02 & 164.65  & 4693.40   & 583.38    \\   \hline
\end{tabular}
\end{ruledtabular}
\end{table}

\section{Nuclear structure and recoil corrections}

When calculating various corrections in the hyperfine structure of the spectrum, 
it is important to note the essential role of corrections for the structure of the 
Li, Be, and B nuclei. Such corrections are determined by the electromagnetic form factors 
of the nuclei. Among the nuclei that we are considering, several nuclei have a spin $s_2=3/2$. 
The amplitude of the one-photon interaction of such nuclei with a muon can be written in the form
\cite{spin321,spin322,spin323}:
\begin{equation}
\label{eq:8}
iM_{1\gamma}=-\frac{Ze^2}{k^2}[\bar u(q_1)\gamma_\mu u(p_1)]
[\bar v_\alpha(p_2){\cal O}_{\alpha\mu\beta}v_\beta(q_2)]=
\end{equation}
\begin{displaymath}
-\frac{Ze^2}{k^2}[\bar u(q_1)\gamma_\mu u(p_1)]
\bar v_\alpha(p_2)\bigl\{g_{\alpha\beta}\frac{(p_2+q_2)_\mu}{2m_2}F_1(k^2)-
g_{\alpha\beta}\sigma_{\mu\nu}\frac{k^\nu}{2m_2}F_2(k^2)+
\end{displaymath}
\begin{displaymath}
\frac{k_\alpha k_\beta}{4m_2^2}\frac{(p_2+q_2)_\mu}{2m_2}F_3(k^2)-
\frac{k_\alpha k_\beta}{4m_2^2}\sigma_{\mu\nu}\frac{k^\nu}{2m_2}F_4(k^2)\bigr\}v_\beta(q_2),
\end{displaymath}
where $p_1$, $p_2$ are four-momenta of particles in the initial state,
$q_1$, $q_2$ are four-momenta of particles in the final state, $k=q_2-p_2=p_1-q_1$.
${\cal O}_{\alpha\mu\beta}$ is the vertex function of the spin 3/2 nucleus.
Nuclei with a spin 3/2 are described by the spin-vector $v_\alpha(p)$. Four form factors $F_i(k^2)$ 
are related to the charge $G_{E0}$, electroquadrupole $G_{E2}$, magnetic dipole $G_{M1}$ 
and magnetic octupole $G_{M3}$ form factors by the following expressions \cite{spin321,spin322,spin323}:
\begin{equation}
\label{eq:9}
G_{E0}=\left(1+\frac{2}{3}\tau\right)[F_1+\tau(F_1-F_2)]-
\frac{\tau}{3}(1+\tau)[F_3+\tau(F_3-F_4)],
\end{equation}
\begin{displaymath}
G_{E2}=F_1+\tau(F_1-F_2)-\frac{1+\tau}{2}[F_3+\tau(F_3-F_4)],
\end{displaymath}
\begin{displaymath}
G_{M1}=(1+\frac{4}{3}\tau)F_2-\frac{2}{3}\tau(1+\tau)F_4,
\end{displaymath}
\begin{displaymath}
G_{M3}=F_2-\frac{1}{2}(1+\tau)F_4, ~~~\tau=-\frac{k^2}{4m_2^2}.
\end{displaymath}

It is useful to consider how the magnitude of the hyperfine splitting in the leading order 
(the Fermi energy)  can be obtained from the amplitude $M_{1\gamma}$. When two moments are added, 
two states appear with the total angular momentum $F = 2$ and $F = 1$. To distinguish the contribution of the amplitude $M_{1\gamma}$ to the interaction operator of particles 
with $F = 2$ and $F=1$, we use the method of projection operators, which are 
constructed from the wave functions of free particles in the rest frame \cite{epja2018,apm1}. 
Thus, the projection operator on a state with $F = 2$ is equal to
\begin{equation}
\label{eq:10}
\hat\Pi_\alpha=[u(0)\bar v_\alpha]_{F=2}=\frac{1+\gamma_0}{2\sqrt{2}}\gamma_\beta\varepsilon_{\alpha\beta},
\end{equation}
where the tensor $\varepsilon_{\alpha\beta}$ describes a muonic atom with F = 2. 
As a result, the projection of $M_{1\gamma}$ to the state with F = 2 takes the form:
\begin{equation}
\label{eq:11}
iM_{1\gamma}(F=2)=-\frac{Ze^2}{16k^2m_1^2m_2^2}Tr\Biggl\{(\hat q_1+m_1)\gamma_\mu(\hat p_1+m_1)
\frac{1+\hat v}{2\sqrt{2}}\gamma_\rho\varepsilon_{\alpha\rho}(\hat p_2-m_2)\times
\end{equation}
\begin{displaymath}
\Bigl[g_{\alpha\beta}\frac{(p_2+q_2)_\mu}{2m_2}F_1(k^2)-
g_{\alpha\beta}\sigma_{\mu\nu}\frac{k^\nu}{2m_2}F_2(k^2)+
\frac{k_\alpha k_\beta}{4m_2^2}\frac{(p_2+q_2)_\mu}{2m_2}F_3(k^2)-
\frac{k_\alpha k_\beta}{4m_2^2}\sigma_{\mu\nu}\frac{k^\nu}{2m_2}F_4(k^2)\Bigr]\times
\end{displaymath}
\begin{displaymath}
(\hat q_2-m_2)\gamma_\lambda
\frac{1+\hat v}{2\sqrt{2}}\varepsilon^\ast_{\beta\lambda}\Biggr\},
\end{displaymath}
where auxiliary four-vector $v=(1,0,0,0)$.
For further construction of the particle interaction potential from~\eqref{eq:11}, 
we use the averaging over the projections of the total angular momentum F which is 
connected with the calculation of the following sum:
\begin{equation}
\label{eq:12}
\sum_{pol}\varepsilon^\ast_{\beta\lambda}\varepsilon_{\alpha\rho}=\hat\Pi_{\beta\lambda\alpha\rho}=
\frac{1}{2}X_{\beta\alpha}X_{\lambda\rho}+\frac{1}{2}X_{\beta\rho}X_{\lambda\alpha}-
\frac{1}{3}X_{\beta\lambda}X_{\alpha\rho},~~~X_{\beta\alpha}=(g_{\alpha\beta}-v_\beta v_\alpha).
\end{equation}
To introduce the projection operators for another state of hyperfine structure
with $F=1$ we use the following expansion:
\begin{equation}
\label{eq:13}
\Psi_{s_2=3/2,F=1,F_z}=\sqrt{\frac{2}{3}}\Psi_{S=0,F=1,F_z}+\frac{1}{\sqrt{3}}\Psi_{S=1,F=1,F_z},
\end{equation}
where the Rarita-Schwinger spinor $v_\alpha(p)$ for the
state with $s_2=3/2$ is presented as a result of adding spin 1/2 and angular 
momentum 1. With this method of adding moments, the total spin S can take
two values $S=1$ and $S=0$. 
When calculating the matrix elements for the states $\Psi_{01F_z}$ and $\Psi_{11F_z}$, 
we successively perform the projection on the state with spin $S=0$, $S=1$, and then 
on the state with the total angular momentum $F = 1$. The corresponding 
projection operators have the form:
\begin{equation}
\label{eq:14}
\hat\Pi_\alpha(S=0,F=1)=\frac{1+\hat v}{2\sqrt{2}}\gamma_5\varepsilon_\alpha,
\end{equation}
\begin{equation}
\label{eq:15}
\hat\Pi_\alpha(S=1,F=1)=\frac{1+\hat v}{4}\gamma_\sigma\varepsilon_{\alpha\sigma\rho\omega}v^\rho\varepsilon^\omega,
\end{equation}
where $\varepsilon^\omega$ is the polarization vector of the state with F=1.
After using \eqref{eq:14}-\eqref{eq:15}, the matrix elements of $M_{1\gamma}$ according 
to the states of $\Psi_{01F_z}$ and $\Psi_{11F_z}$ are reduced to the form:
\begin{equation}
\label{eq:16}
<\Psi_{01F_z}|iM_{1\gamma}(F=1)|\Psi_{01F_z}>=\frac{\pi Z\alpha}{96 k^2m_1^2m_2^2}
Tr\Bigl\{(\hat q_1+m_1)\gamma_\mu(\hat p_1+m_1)(1+\hat v)\gamma_5(\hat p_2-m_2)\times
\end{equation}
\begin{displaymath}
\Bigl[g_{\alpha\beta}\frac{(p_2+q_2)_\mu}{2m_2}F_1(k^2)-
g_{\alpha\beta}\sigma_{\mu\nu}\frac{k^\nu}{2m_2}F_2(k^2)+
\frac{k_\alpha k_\beta}{4m_2^2}\frac{(p_2+q_2)_\mu}{2m_2}F_3(k^2)-
\frac{k_\alpha k_\beta}{4m_2^2}\sigma_{\mu\nu}\frac{k^\nu}{2m_2}F_4(k^2)\Bigr]
\end{displaymath}
\begin{displaymath}
(\hat q_2-m_2)\gamma_5(1+\hat v)\Bigr\}(-g_{\alpha\beta}+v_\alpha v_\beta),
\end{displaymath}
\begin{equation}
\label{eq:17}
<\Psi_{11F_z}|iM_{1\gamma}(F=1)|\Psi_{11F_z}>=\frac{\pi Z\alpha}{192 k^2m_1^2m_2^2}
Tr\Bigl\{(\hat q_1+m_1)\gamma_\mu(\hat p_1+m_1)(1+\hat v)\gamma_\sigma(\hat p_2-m_2)\times
\end{equation}
\begin{displaymath}
\Bigl[g_{\alpha\beta}\frac{(p_2+q_2)_\mu}{2m_2}F_1(k^2)-
g_{\alpha\beta}\sigma_{\mu\nu}\frac{k^\nu}{2m_2}F_2(k^2)+
\frac{k_\alpha k_\beta}{4m_2^2}\frac{(p_2+q_2)_\mu}{2m_2}F_3(k^2)-
\frac{k_\alpha k_\beta}{4m_2^2}\sigma_{\mu\nu}\frac{k^\nu}{2m_2}F_4(k^2)\Bigr]
\end{displaymath}
\begin{displaymath}
(\hat q_2-m_2)\gamma_\epsilon(1+\hat v)\Bigr\}
\varepsilon_{\alpha\sigma\rho\omega}\varepsilon{\beta\epsilon\tau\lambda}
(-g_{\lambda\omega}+v_\lambda v_\omega).
\end{displaymath}

\begin{table}[htbp]
\caption{\label{tb3}Hyperfine splittings of S-states in muonic
ion $(\mu~^ 9_4Be)^{3+}$.}
\bigskip
\begin{ruledtabular}
\begin{tabular}{|c|c|c|c|}  \hline
 No. & Contribution to the splitting &\multicolumn{2}{c|}{$(\mu~^ 9_4Be)^{3+}$,~meV} \\   \hline
  & &1S~~~~~~~ & 2S \\   \hline
1 & Contribution of order $\alpha^4$, &  -4353.49   & -544.19   \\ 
  &  the Fermi energy &    &       \\    \hline
2 & Muon AMM contribution  &  -5.08   &  -0.64    \\  \hline
3 & Relativistic correction &   -5.57    &   -0.99      \\
  & of order $\alpha^6$   &           &                         \\   \hline
4 & Nuclear structure correction & G;~441.09 & G:~55.14    \\ 
  &         of order $\alpha^5$                  & U:~449.54 &U:~56.19  \\  \hline
5 & Nuclear structure and recoil & G:~-97.71 & G:~-12.21  \\  \hline  
6 &  Nuclear structure correction  &  -17.57  &   -1.78      \\
 & of order $\alpha^6$ in  $1\gamma$ interaction    &     &           \\   \hline
7 & Nuclear structure correction in second & 12.64  &   4.36       \\
 & order perturbation theory    &     &             \\   \hline
8 & Vacuum polarization contribution & -17.97&   -2.30     \\
 & of order $\alpha^5$ in first order PT &       &             \\   \hline
9 & Vacuum polarization contribution &-42.62 & -3.92       \\
 & of order $\alpha^5$ in second order PT &       &              \\   \hline
10 & Muon vacuum polarization contribution &-0.34 & -0.04        \\
 & of order $\alpha^6$ in first order PT &       &           \\   \hline
11 & Muon vacuum polarization contribution &-0.36 &-0.05         \\
 & of order $\alpha^6$ in srcond order PT &       &           \\   \hline
12 & Vacuum polarization contribution   & -0.24 &  -0.03    \\
 & of order $\alpha^6$ in first order PT  &      &          \\   \hline
13& Vacuum polarization contribution  & -0.54 & -0.05      \\
 &  of order $\alpha^6$ in second order PT         &   &           \\   \hline
14 & Nuclear structure and vacuum & 5.31  &  0.66    \\
 &  polarization correction of order  $\alpha^6$ &     &      \\    \hline
15& Nuclear structure and muon vacuum  &0.55   &  0.07        \\
 & polarization correction of order  $\alpha^6$ &     &        \\    \hline
16 & Hadron vacuum polarization  &   -0.25  &   -0.03     \\
 & contribution of order $\alpha^6$   &     &                  \\    \hline
17& Radiative nuclear finite size  & 1.44   &  0.18     \\
 &  correction of order  $\alpha^6$    &     &            \\   \hline
&  Summary contribution  & -4080.71 &   -505.82    \\   \hline
\end{tabular}
\end{ruledtabular}
\end{table}

In addition, the off-diagonal matrix element $<\Psi_{01F_z}|iM_{1\gamma}(F=1)|\Psi_{11F_z}>$
is also nonzero. The sum of all the matrix elements gives, in the nonrelativistic approximation, 
the following value of the hyperfine splitting (the Fermi energy) (see the numerical values in
Tables~\ref{tb2}-\ref{tb4}, line 1):
\begin{equation}
\label{eq:18}
\Delta E^{hfs}_{1\gamma}=E_F(nS)=\frac{16}{9}\frac{\pi Z\alpha}{m_1m_2}F_2(0)\frac{W^3}{\pi n^3}=
\frac{16\alpha(Z\alpha)^3\mu^3}{9m_1m_p n^3}\mu_N.
\end{equation}

\begin{figure}[htbp]
\centering
\includegraphics[scale=0.8]{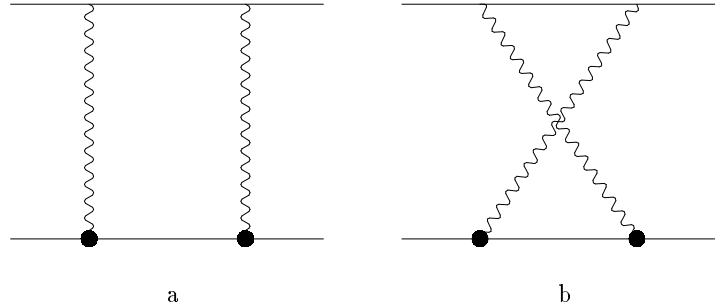}
\caption{Nuclear structure effects of order $\alpha^5$. The bold point denotes the nucleus
vertex function.}
\label{fig:pic3}
\end{figure}

Expressions \eqref{eq:16}-\eqref{eq:17} are presented in a form that is convenient 
for the subsequent calculation of the contribution in the Form package \cite{form}.
We present in detail the results of calculating the amplitude $M_{1\gamma}$, since this 
calculation technique is used later in the calculation of two-photon exchange amplitudes.
In the case of nuclei with spin $s_2=1/2$ and $s_2=1$ the similar technique of projection
operators was used in \cite{apm1,apm2,apm3,apm4}.

\begin{figure}[htbp]
\centering
\includegraphics{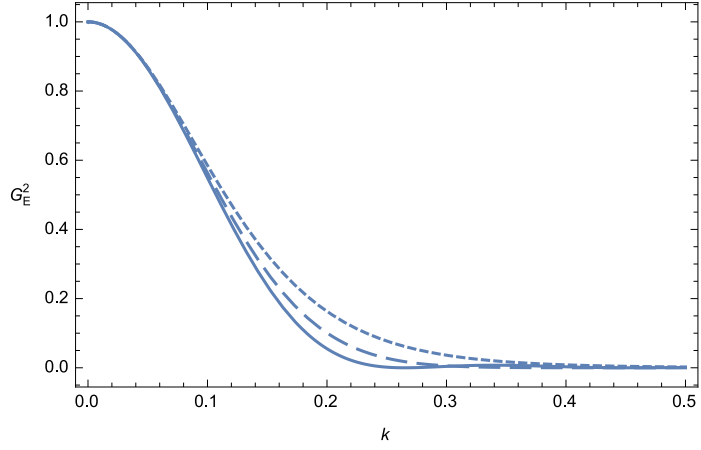}
\caption{Gaussian (dashed), dipole (dotted) and uniformly charged sphere (solid) parameterizations of nuclear form factor $G_E^2(k^2)$.}
\label{fig:ff}
\end{figure}

Basic contribution of the nuclear structure effects of order $\alpha^5$ to
the hyperfine splitting is determined by two-photon exchange diagrams shown in
Fig.~\ref{fig:pic3}. It is expressed in terms of electric $G_E(k^2)$ and
magnetic $G_M(k^2)$ nuclear form factors in the form (the Zemach correction):
\begin{equation}
\label{eq:19}
\Delta E^{hfs}_{str}=E_F\frac{2\mu Z\alpha}{\pi}\int\frac{d{\bf k}}{{\bf k}^4}
\left[\frac{G_E(k^2)G_M(k^2)}{G_M(0)}-1\right].
\end{equation}
We have analysed numerical values of correction \eqref{eq:19} for different
parameterizations of nuclear form factors: Gaussian $G_E^G(k^2)$, dipole $G_E^D(k^2)$
and uniformly charged sphere $G_E^U(k^2)$:
\begin{equation}
\label{eq:20}
G_E^G(k^2)=e^{-\frac{1}{6}r_N^2k^2},~~~
G_E^D(k^2)=\frac{1}{(1+\frac{k^2}{\Lambda^2})^2},~~~
G_E^U(k^2)=\frac{3}{(kR)^3}\left[\sin kR-kR\cos kR\right],
\end{equation}
where $R=\sqrt{5}r_N/\sqrt{3}$ is the nucleus radius, $\Lambda^2=12/r_N^2$. A comparison of functions $G_E^2(k^2)$
for different parameterizations is presented in Fig.~\ref{fig:ff} for the nucleus $^6_3Li$. In the range 
$0.1\leq k\leq 0.4$ GeV there is a difference between functions \eqref{eq:19} which leads 
to different numerical values of the Zemach correction shown in Tables~\ref{tb2}-\ref{tb4} (line 4).

The momentum integration in \eqref{eq:19} can be done analytically, so that the
Zemach correction with the Gaussian and uniformly charged sphere parameterizations has
the form (numerical results are presented in Tables~\ref{tb2}-\ref{tb4} for these two parametrizations):
\begin{equation}
\label{eq:21}
\Delta E^{hfs}_{str,G}=-E_F\frac{72}{\sqrt{3\pi}}\mu Z\alpha r_N,~~~
\Delta E^{hfs}_{str,U}=-E_F\frac{72\sqrt{5}}{35\sqrt{3}}\mu Z\alpha r_N.
\end{equation}

\begin{table}[htbp]
\caption{\label{tb4}Hyperfine splittings of S-states in muonic
ions $(\mu~^ {10}_5B)^{4+}$ and $(\mu~^ {11}_5B)^{4+}$.}
\bigskip
\begin{ruledtabular}
\begin{tabular}{|c|c|c|c|c|c|}  \hline
 No. & Contribution to the splitting &\multicolumn{2}{c|}{$(\mu~^ {10}_5B)^{4+}$,~meV}&
\multicolumn{2}{c|}{$(\mu~^ {11}_5B)^{4+}$, meV }   \\   \hline
  & &1S~~~~~~~ & 2S & 1S~~~~~~~& 2S \\   \hline
1 & Contribution of order $\alpha^4$, & 11420.56    &1427.57 & 19548.21& 2443.53  \\ 
  & the Fermi energy   &    &    &     &    \\    \hline
2 & Muon AMM contribution  & 13.33    & 1.67& 22.82 & 2.85     \\  \hline
3 & Relativistic correction &  22.83     & 4.04 &39.08 &  6.92     \\
  & of order $\alpha^6$   &           &              &      &           \\   \hline
4 & Nuclear structure correction & G;~-1395.72 & G:~-174.46&G:~-2370.05&G:~-296.26 \\ 
  &         of order $\alpha^5$        & U:~-1422.43 &U:~-177.80 &U:~-2415.42 & U:~-301.93  \\  \hline
5 & Nuclear structure and recoil & --- & ---&G:~36.68&G:~4.59 \\  \hline
6 &  Nuclear structure correction  & 67.06   & 6.81    &112.97   & 11.47    \\
 & of order $\alpha^6$ in  $1\gamma$ interaction    &     &   &  &         \\   \hline
7 & Nuclear structure correction in second & -46.40  & -15.69 & -78.15& -26.45         \\
 & order perturbation theory    &     &      &    &        \\   \hline
8 & Vacuum polarization contribution & 50.99& 6.51 & 87.31 &   11.14    \\
 & of order $\alpha^5$ in first order PT &       &   &   &            \\   \hline
9 & Vacuum polarization contribution & 123.21&11.38  & 210.99 &19.49       \\
 & of order $\alpha^5$ in second order PT &       &   &   &            \\   \hline
10 & Muon vacuum polarization contribution &1.10 &0.14  &1.89  &  0.24     \\
 & of order $\alpha^6$ in first order PT &       &   &   &            \\   \hline
11 & Muon vacuum polarization contribution &1.21 &0.15  &2.07  &  0.26     \\
 & of order $\alpha^6$ in second order PT &       &   &   &            \\   \hline
12 & Vacuum polarization contribution   & 0.71 &0.09  &1.21  &  0.15    \\
 & of order $\alpha^6$ in first order PT  &      &      &   &      \\   \hline
13 & Vacuum polarization contribution  & 1.63 & 0.15 &2.79   &  0.27   \\
 &  of order $\alpha^6$ in second order PT         &   &      &       &      \\   \hline
14 & Nuclear structure and vacuum &  -20.80 & -2.60  &-35.06  & -4.38    \\
 &  polarization correction of order  $\alpha^6$ &     &    &   &    \\    \hline
15& Nuclear structure and muon vacuum  & -1.90  & -0.24  &-3.26 &  -0.41       \\
 & polarization correction of order  $\alpha^6$ &     &  &   &        \\    \hline
16 & Hadron vacuum polarization  &  0.81   & 0.10  & 1.38 &  0.17     \\
 &contribution  of order $\alpha^6$   &     &   &       &                  \\    \hline
17& Radiative nuclear finite size  &  -4.76  & -0.59  & -8.18& -1.02      \\
 &  correction of order  $\alpha^6$    &     &   &   &          \\   \hline
&  Summary contribution  & 10233.86 & 1265.03  &  17572.70  & 2172.56    \\   \hline
\end{tabular}
\end{ruledtabular}
\end{table}

Acting as in the case of the one-photon interaction, we can present the contribution
of two-photon interactions to HFS at $F=2$ in the form:
\begin{equation}
\label{eq:22}
\Delta E^{hfs}_{2\gamma}(F=2)=-\frac{(Z\alpha)^2}{640\pi^2m_1^2m_2^2}|\psi(0)|^2\int
\frac{id^4k(k^2-2k_0m_2)}{k^4(k^4-4k_0^2m_1^2)(k^4-4k_0^2m_2^2)}Tr\bigl\{
(\hat q_1+m_1)\times
\end{equation}
\begin{displaymath}
[\gamma_\mu(\hat p_1-\hat k+m_1)
\gamma_\nu(k^2+2k_0m_1)+\gamma_\nu(\hat p_1+\hat k+m_1)\gamma_\mu(k^2-2k_0m_1)](\hat p_1+m_1)
(1+\hat v)\gamma_\rho(\hat p_2-m_2)\times
\end{displaymath}
\begin{displaymath}
{\cal O}_{\alpha\nu\sigma}(k)
(-\hat p_2-\hat k+m_2)[g_{\sigma\tau}-\frac{1}{3}\gamma_\sigma\gamma_\tau-
\frac{2}{3m_2^2}(p_2+k)_\sigma(p_2+k)_\tau+\frac{1}{3m_2}(\gamma_\sigma(p_2+k)_\tau-
\gamma_\tau(p_2+k)_\sigma)]\times
\end{displaymath}
\begin{displaymath}
{\cal O}_{\tau\mu\beta}(-k)(\hat q_2-m_2)\gamma_\lambda(1+\hat v)\bigr\}
\hat\Pi_{\beta\alpha\lambda\rho},
\end{displaymath}
where $k$ is a loop momentum, and $k_0$ its zero component.
We also give for completeness analogous expressions for two states in \eqref{eq:13} 
with $F=1, S=0$ and $F=1, S=1$:
\begin{equation}
\label{eq:23}
\Delta E^{hfs}_{2\gamma}(F=1,S=0)=-\frac{(Z\alpha)^2}{384\pi^2m_1^2m_2^2}|\psi(0)|^2\int
\frac{id^4k(k^2-2k_0m_2)}{k^4(k^4-4k_0^2m_1^2)(k^4-4k_0^2m_2^2)}Tr\bigl\{
(\hat q_1+m_1)\times
\end{equation}
\begin{displaymath}
[\gamma_\mu(\hat p_1-\hat k+m_1)
\gamma_\nu(k^2+2k_0m_1)+\gamma_\nu(\hat p_1+\hat k+m_1)\gamma_\mu(k^2-2k_0m_1)](\hat p_1+m_1)
(1+\hat v)\gamma_5(\hat p_2-m_2)\times
\end{displaymath}
\begin{displaymath}
{\cal O}_{\alpha\nu\sigma}(k)
(-\hat p_2-\hat k+m_2)[g_{\sigma\tau}-\frac{1}{3}\gamma_\sigma\gamma_\tau-
\frac{2}{3m_2^2}(p_2+k)_\sigma(p_2+k)_\tau+\frac{1}{3m_2}(\gamma_\sigma(p_2+k)_\tau-
\gamma_\tau(p_2+k)_\sigma)]\times
\end{displaymath}
\begin{displaymath}
{\cal O}_{\tau\mu\beta}(-k)(\hat q_2-m_2)\gamma_5(1+\hat v)\bigr\}
(-g_{\alpha\beta}+v_\alpha v_\beta),
\end{displaymath}
\begin{equation}
\label{eq:24}
\Delta E^{hfs}_{2\gamma}(F=1,S=1)=-\frac{(Z\alpha)^2}{768\pi^2m_1^2m_2^2}|\psi(0)|^2\int
\frac{id^4k(k^2-2k_0m_2)}{k^4(k^4-4k_0^2m_1^2)(k^4-4k_0^2m_2^2)}Tr\bigl\{
(\hat q_1+m_1)\times
\end{equation}
\begin{displaymath}
[\gamma_\mu(\hat p_1-\hat k+m_1)
\gamma_\nu(k^2+2k_0m_1)+\gamma_\nu(\hat p_1+\hat k+m_1)\gamma_\mu(k^2-2k_0m_1)](\hat p_1+m_1)
(1+\hat v)\gamma_{\sigma_1}\varepsilon_{\alpha\sigma_1\rho_1\omega_1}v_{\rho_1}
\times
\end{displaymath}
\begin{displaymath}
(\hat p_2-m_2)
{\cal O}_{\alpha\nu\sigma}(k)
(-\hat p_2-\hat k+m_2)[g_{\sigma\tau}-\frac{1}{3}\gamma_\sigma\gamma_\tau-
\frac{2}{3m_2^2}(p_2+k)_\sigma(p_2+k)_\tau+\frac{1}{3m_2}(\gamma_\sigma(p_2+k)_\tau-
\end{displaymath}
\begin{displaymath}
\gamma_\tau(p_2+k)_\sigma)]
{\cal O}_{\tau\mu\beta}(-k)(\hat q_2-m_2)\gamma_{\epsilon_1}(1+\hat v)\varepsilon_{\beta\epsilon_1\tau_1\lambda_1}v_{\tau_1}\bigr\}
(-g_{\lambda_1\omega_1}+v_{\lambda_1} v_{\omega_1}).
\end{displaymath}
There is also an off-diagonal matrix element between states
with $F=1, S=0$ and $F=1, S=1$, which we omit here.
The expressions \eqref{eq:21} - \eqref{eq:24} are presented in a form convenient for the subsequent calculation in the package Form \cite{form}.
As a result the value of the hyperfine splitting is determined in Euclidean space by
the following formula:
\begin{equation}
\label{eq:25}
\Delta E^{hfs}(nS)=|\psi_{nS}(0)|^2\int d^4k V_{2\gamma}(k)=\frac{64}{9}\frac{(Z\alpha)^2}{\pi^2}|\psi_{nS}(0)|^2\int \frac{d^4k}
{k^4(k^4+4m_1^2k_0^2)(k^4+4m_2^2k_0^2)}\times
\end{equation}
\begin{displaymath}
\Bigl[F_1F_2\Bigl(k^6-k^4k_0^2+\frac{4}{15}\frac{k^4k_0^4}{m_2^2}-\frac{7}{10}
\frac{k^6k_0^2}{m_2^2}+\frac{13}{30}\frac{k^8}{m_2^2}\Bigr)+
F_2F_4\Bigl(-\frac{1}{30}\frac{k^2k_0^6}{m_2^2}+\frac{1}{15}
\frac{k^4k_0^4}{m_2^2}-\frac{1}{30}\frac{k^6k_0^2}{m_2^2}\Bigr)+
\end{displaymath}
\begin{displaymath}
F_2F_3\Bigl(-\frac{1}{15}\frac{k^2k_0^6}{m_2^2}+\frac{11}{60}
\frac{k^4k_0^4}{m_2^2}-\frac{7}{60}\frac{k^8}{m_2^2}\Bigr)+
F_1F_4\Bigl(-\frac{1}{5}\frac{k^2k_0^6}{m_2^2}+\frac{3}{10}
\frac{k^4k_0^4}{m_2^2}-\frac{1}{10}\frac{k^8}{m_2^2}\Bigr)+
\end{displaymath}
\begin{displaymath}
F_2^2\Bigl(\frac{1}{15}\frac{k^2k_0^6}{m_2^2}-\frac{1}{6}k^2k_0^4-\frac{2}{15}
\frac{k^4k_0^4}{m_2^2}+\frac{1}{6}k^4k_0^2+\frac{23}{120}\frac{k^6k_0^2}{m_2^2}
-\frac{1}{4}\frac{k^8}{m_2^2}\Bigr)\Bigr].
\end{displaymath}
When investigating this expression, it is useful to distinguish the Zemach correction, 
which is determined by the integral
\begin{equation}
\label{eq:26}
J=\int_0^\infty\int_0^\pi\frac{k\sin^2\phi dk d\phi F_1(k^2)F_2(k^2)}{(k^2+4m_1^2\cos^2\phi)
(k^2+4m_2^2\cos^2\phi)}
=\frac{\pi}{2(m_1+m_2)}\int_0^\infty\frac{ dk }{k^2}F_1(k^2)F_2(k^2)+
\end{equation}
\begin{displaymath}
\frac{\pi}{4(m_1^2-m_2^2)}\int_0^\infty\frac{dk}{k^2}[\sqrt{k^2+4m_1^2}-2m_1-
\sqrt{k^2+4m_2^2}+2m_2]F_1(k^2)F_2(k^2).
\end{displaymath}
The divergence in the first term on the right-hand side of \eqref{eq:26} is 
compensated by the subtraction term
\begin{equation}
\label{eq:27}
\Delta E^{hfs}_{iter}=\frac{64}{9}\frac{(Z\alpha)^2}{\pi^2}|\psi(0)|^2\int_0^\infty
\frac{2\pi^2F_2(0)}{(m_1+m_2)k^2}dk.
\end{equation}
Thus, we have in \eqref{eq:25} the main contribution (the Zemach correction) and the recoil correction $m_1/m_2$.
In the case of Li nucleus with spin 1 the expression similar to \eqref{eq:25} was
obtained in \cite{apm3} for muonic deuterium. We use it for corresponding numerical
estimates of nuclear structure and recoil corrections in $^6_3Li$.
The recoil effects for the nucleus $^{10}_5B$ with spin 3 are neglected.

The form factors $F_i(k^2)$ are expressed in terms of $G_{E0}$,  $G_{E2}$, 
$G_{M1}$, $G_{M3}$ for which the 
Gaussian parametrization is used in numerical calculations of integrals with respect to k.
The values of the form factors at zero have the form:
\begin{equation}
\label{eq:28}
G_{E0}(0)=1,~~~G_{M1}(0)=\frac{m_2\mu_N}{m_p Z},~~~G_{E2}(0)=m_2^2Q,~~~G_{M3}(0)=
\frac{m_2}{m_pZ}m_2^2\Omega.
\end{equation}
Different parameters of light nucleus (Li, Be, B) were investigated in
electron scattering experiments \cite{scat,fuller}.
The nucleus multiple moments are presented in Table~\ref{tb1}.
Some of them are unknown with good accuracy, but, nevertheless, 
one can obtain approximate estimates of the corresponding contributions.
After angular analytical integration in \eqref{eq:25} we make numerical integration
over $k$. Obtained results for nuclear structure and recoil corrections are
presented in Tables~\ref{tb2}-\ref{tb4} in separate lines.

\begin{figure}[htbp]
\centering
\includegraphics[scale=1.0]{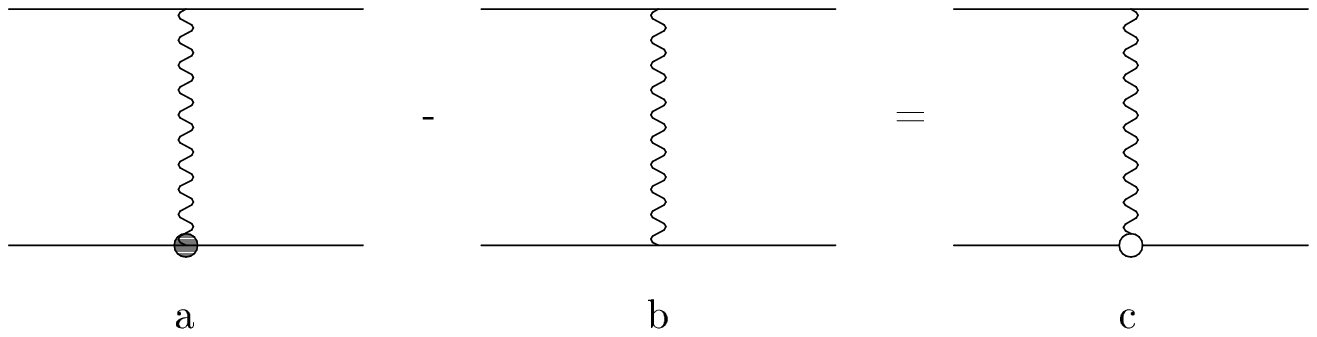}\\[5mm]
\includegraphics[scale=0.8]{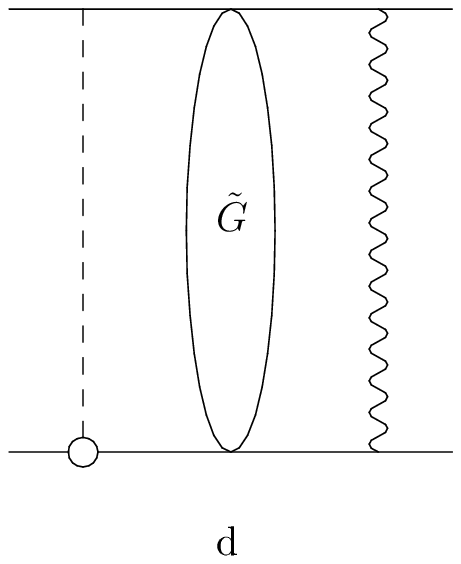}
\caption{Nuclear structure effects in one-photon interaction (c) and in second order perturbation
theory (d). $\tilde G$ is the reduced Coulomb Green's function.}
\label{fig:pic5}
\end{figure}

Another correction for the structure of the nucleus of order $\alpha^6$, which must be discussed, 
is obtained as a result of the decomposition of the magnetic form factor 
of the nucleus see Fig.~\ref{fig:pic5}(a)). 
The contribution to the interaction potential and HFS in this case has the form
\cite{apm2008,apm2}:
\begin{equation}
\label{eq:28a}
\Delta V^{hfs}_{1\gamma,str}(r)=\frac{4\pi\alpha\mu_N}{9m_1m_p}r_M^2({\bf s}_1{\bf s}_2)
\nabla^2\delta({\bf r}),
\end{equation}
\begin{equation}
\label{eq:28b}
\Delta E^{hfs}_{1\gamma,str}=\frac{2}{3}\mu^2Z^2\alpha^2r_M^2\frac{3n^2+1}{n^2}E_F(nS).
\end{equation}
Numerical values on the basis \eqref{eq:28b} can be obtained assuming that $r_M^2=r_E^2$.
They are in line 11 of Tables~\ref{tb2}-\ref{tb4}.

In second order PT we should take into account a term in which the potential
\begin{equation}
\label{eq:28c}
\Delta V^{C}_{str,1\gamma}(k)=-\frac{Z\alpha}{{\bf k}^2}
\left[\frac{3}{(kR)^3}\left(\sin kR-kR\cos kR\right)-1\right]
\end{equation}
is considered as a perturbation. The Fourier transform of \eqref{eq:28c} is
\begin{equation}
\label{eq:28d}
\Delta V^C_{str,1\gamma}(r)=-\frac{Z\alpha}{4R^3r}(r-R)(r+2R)(R-r+|r-R|).
\end{equation}
Using the Green's functions~\eqref{eq:37} and \eqref{eq:38} (see section III) we perform the analytical integration 
in second order PT. It gives the following result:
\begin{equation}
\label{eq:28e}
\Delta E^{hfs}_{str,sopt}(1S)=-E_F(1S)\frac{R^2W^2}{4}\bigl[
-\frac{4}{75}(-53+15C+15\ln RW+\frac{RW}{12}\bigl(-15+4C+4\ln RW\bigr)\bigl],
\end{equation}
\begin{equation}
\label{eq:28f}
\Delta E^{hfs}_{str,sopt}(2S)=E_F(2S)\frac{R^2W^2}{4}\bigl[
\frac{4}{75}(-107+60C+60\ln RW)+\frac{RW}{3}(17-8C-8\ln RW)
\bigr],
\end{equation}
where we present an expansions in $(RW)$ up to terms of first order in square brackets
($RW(^6_3Li)=0.038$, $RW(^7_3Li)=0.036$, $RW(^9_4Be)=0.050$, $RW(^{10}_5B)=0.060$, $RW(^{11}_5B)=0.060$).
Numerical values of \eqref{eq:28e} and \eqref{eq:28f} are sufficiently important (see line 7
of Tables~\ref{tb2}-\ref{tb4}).

\begin{figure}[htbp]
\centering
\includegraphics{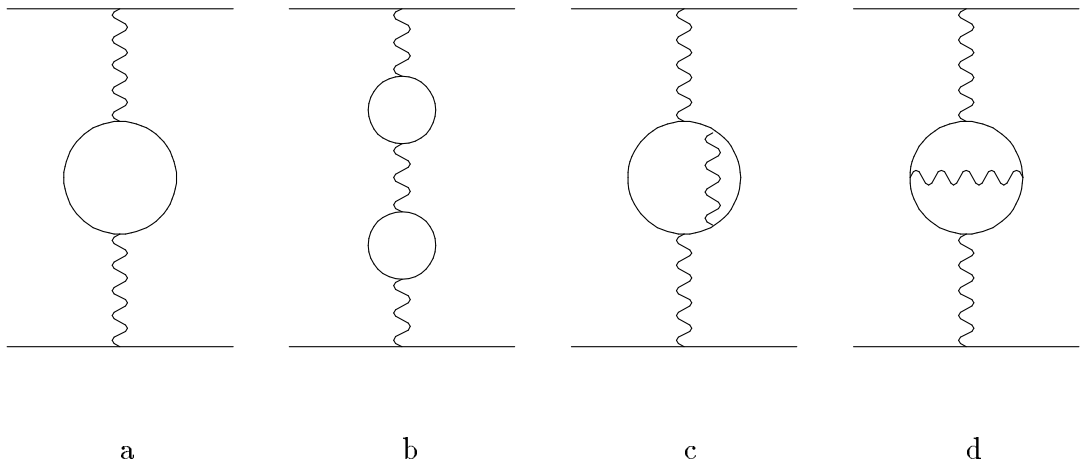}
\caption{Effects of one- and two-loop vacuum polarization in one-photon interaction.}
\label{fig:pic1}
\end{figure}

\section{Effects of one- and two-loop vacuum polarization in first and second 
orders of perturbation theory}

Another our task is to analyse different vacuum polarization corrections to the total value of the hyperfine splitting. Primarily, we have to calculate a contribution of one-loop vacuum polarization contribution in first order PT. Corresponding potential can be obtained
in momentum representation after a standard modification of hyperfine muon-nucleus 
interaction due to vacuum polarization effect \cite{borie1,borie2,kp1996,sgk1,sgk2,t4}.
In coordinate representation it is defined by the following integral expression:
\begin{equation}
\label{eq:29}
\Delta V^{hfs}_{1\gamma,vp}(r)=\frac{4\alpha g_N(1+a_\mu)}{3m_1m_p}
({\bf s}_1{\bf s}_2)
\frac{\alpha}{3\pi}
\int_1^\infty\rho(\xi)d\xi\left(\pi\delta({\bf r})-\frac{m_e^2\xi^2}{r}e^{-2m_e\xi r}\right),
\end{equation}
where $g_N=\mu_N/s_2$, spectral function $\rho(\xi)=\sqrt{\xi^2-1}(2\xi^2+1)/\xi^4$.
We include in \eqref{eq:29} the anomalous magnetic moment of muon, which leads to the 
additional contribution of order $\alpha^6$. Averaging~\eqref{eq:29} over wave 
functions~\eqref{eq:4} and \eqref{eq:5}, we get the
contribution of order $\alpha^5$ to hyperfine structure of $1S-$ and $2S$-states ($a_1=m_e/W$,
$W=\mu Z\alpha$):
\begin{equation}
\label{eq:30}
\Delta E_{1\gamma,vp}^{hfs}(1S)=\frac{4\alpha^2(Z\alpha)^3\mu^3 g_N (1+a_\mu)}{9m_1m_p\pi}
<({\bf s}_1{\bf s}_2)>\int_1^\infty
\rho(\xi)d\xi\left[1-\frac{m_e^2\xi^2}{W^2}\int_0^\infty xdxe^{-x\left(1+\frac{m_e\xi}{W}\right)}\right]=
\end{equation}
\begin{displaymath}
E_F(1S)\frac{\alpha(1+a_\mu)}{9\pi\sqrt{1-a_1^2}}\Bigl[\sqrt{1-a_1^2}(1+6a_1^2-3\pi a_1^3)+
(6-3a_1^2+5a_1^4)\ln\frac{1+\sqrt{1-a_1^2}}{a_1}\Bigr],
\end{displaymath}
\begin{equation}
\label{eq:31}
\Delta E_{1\gamma,vp}^{hfs}(2S)=\frac{\alpha^2(Z\alpha)^3\mu^3g_N(1+a_\mu)}{18m_1m_p\pi}
<({\bf s}_1{\bf s}_2)>\int_1^\infty\rho(\xi)d\xi\times
\end{equation}
\begin{displaymath}
\left[1-\frac{4m_e^2\xi^2}{W^2}\int_0^\infty x\left(1-\frac{x}{2}\right)^2dxe^{-x\left(1+\frac{2m_e\xi}{W}\right)}\right]=
\end{displaymath}
\begin{displaymath}
E_F(2S)\frac{\alpha(1+a_\mu)}{18\pi(4a_1^2-1)^{5/2}}\Bigl\{\sqrt{4a_1^2-1}[11+2a_1^2
(-29+8a_1(-22a_1+48a_13-3\pi(4a_1^2-1)^2))]+
\end{displaymath}
\begin{displaymath}
12(1-10a_1^2+66a_1^4-160a_1^6+256a_1^8)\arctan\sqrt{4a_1^2-1}\Bigr\}.
\end{displaymath}
We present in detail the results of \eqref{eq:30}-\eqref{eq:31} to demonstrate the general structure of the obtained analytical expressions. After integrating over particle coordinates, the results have a fairly simple form, but the following integration over the spectral parameters gives, as a rule, rather cumbersome expressions, which we will omit in the following.

With a simple replacement $m_e$ to muon mass $m_1$ in equations \eqref{eq:30} and 
\eqref{eq:31} one can obtain the muon vacuum polarization correction to HFS of order 
$\alpha^6$. The numerical values of the muon vacuum polarization corrections are included in Table~\ref{tb2}-\ref{tb4} in the corresponding line (line 6). Another contribution of order 
$\alpha^6$ is represented by two-loop vacuum polarization diagrams
(see Fig.~\ref{fig:pic1}(b,c,d) (the K\"allen and Sabry potential \cite{sabry}).
The construction of the interaction potentials from these diagrams is completely 
analogous to \eqref{eq:29}. They have the form of a double and a single spectral integral
in coordinate space \cite{apm2004,apm2008}:
\begin{equation}
\label{eq:32}
\Delta V_{1\gamma,vp-vp}^{hfs}(r)=\frac{4\pi\alpha g_N(1+a_\mu)}{3m_1m_p}
({\bf s}_1{\bf s}_2)\left(\frac{\alpha}
{3\pi}\right)^2\int_1^\infty\rho(\xi)d\xi\int_1^\infty\rho(\eta)d\eta\times
\end{equation}
\begin{displaymath}
\times\left[\delta({\bf r})-\frac{m_e^2}{\pi r(\eta^2-\xi^2)}\left(\eta^4 e^{-2m_e\eta r}-
\xi^4 e^{-2m_e\xi r}\right)\right],
\end{displaymath}
\begin{equation}
\label{eq:33}
\Delta V_{1\gamma,2-loop~vp}^{hfs}(r)=\frac{8\alpha^3g_N(1+a_\mu)}
{9\pi^2 m_1m_p}({\bf s}_1{\bf s}_2)
\int_0^1\frac{f(v)dv}{1-v^2}\left[\pi\delta({\bf r})-\frac{m_e^2}{ r(1-v^2)}e^{-\frac{2m_er}{\sqrt{1-v^2}}}\right],
\end{equation}
where two-loop spectral function
\begin{equation}
\label{eq:34}
f(v)=v\Bigl\{(3-v^2)(1+v^2)\left[Li_2\left(-\frac{1-v}{1+v}\right)+2Li_2
\left(\frac{1-v}{1+v}\right)+\frac{3}{2}\ln\frac{1+v}{1-v}\ln\frac{1+v}{2}-
\ln\frac{1+v}{1-v}\ln v\right]
\end{equation}
\begin{displaymath}
+\left[\frac{11}{16}(3-v^2)(1+v^2)+\frac{v^4}{4}\right]\ln\frac{1+v}{1-v}+
\left[\frac{3}{2}v(3-v^2)\ln\frac{1-v^2}{4}-2v(3-v^2)\ln v\right]+
\frac{3}{8}v(5-3v^2)\Bigr\},
\end{displaymath}
where $Li_2(z)$ is the Euler dilogarithm.
Averaging \eqref{eq:32}-\eqref{eq:33} over wave functions \eqref{eq:5}-\eqref{eq:6}
the integration over $r$ can be done analytically, while two other integrations 
over $\xi$ and $\eta$ are calculated numerically with the use of Wolfram Mathematica.
Summary two-loop vacuum polarization correction of order $\alpha^6$ is written in
Tables~\ref{tb2}-\ref{tb4} (line 7).

\begin{figure}[htbp]
\centering
\includegraphics{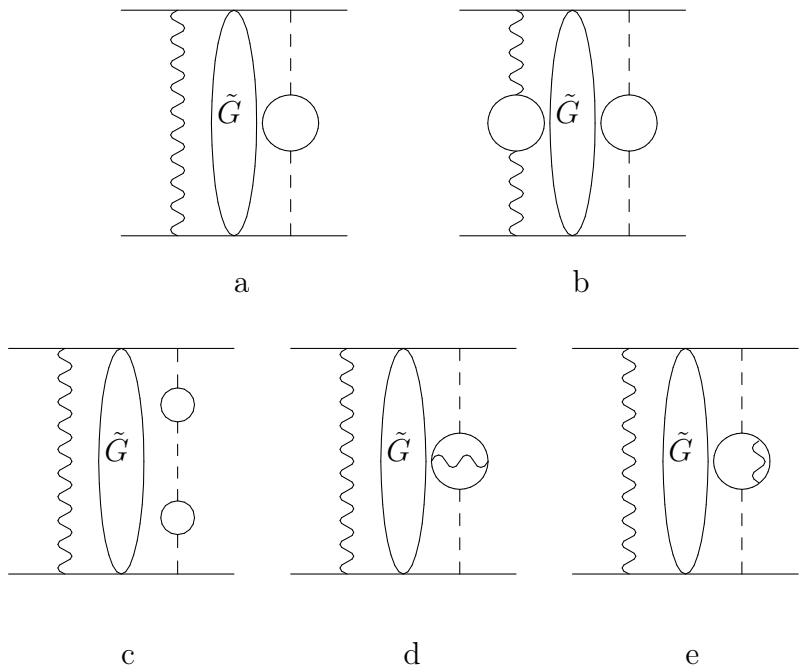}
\caption{Effects of one- and two-loop vacuum polarization in second order PT.}
\label{fig:pic2}
\end{figure}

To achieve the desired accuracy of calculations one-loop and two-loop contributions of order $\alpha^5$ and $\alpha^6$ to HFS have to be taken into account in second order perturbation theory.
The second order perturbation theory (PT) corrections to the energy spectrum are
determined by the reduced Coulomb Green's function $\tilde G_n({\bf r}, {\bf r'})$
The radial part $\tilde g_{nl}(r,r')$ of $\tilde G_n({\bf r}, {\bf r'})$
was obtained in~\cite{hameka} in the form of the Sturm
expansion in the Laguerre polynomials. The main contribution of the electron vacuum polarization
to HFS in second order PT (sopt) has the form (see Fig.~\ref{fig:pic2}(a)):
\begin{equation}
\label{eq:35}
\Delta E^{hfs}_{sopt~vp~1}=2<\psi|\Delta V^C_{VP}\cdot \tilde G\cdot\Delta
V_B^{hfs}|\psi>,
\end{equation}
where the Coulomb potential, modified by the one-loop vacuum polarization effect, has the form
\begin{equation}
\label{eq:36}
\Delta V^C_{vp}(r)=\frac{\alpha}{3\pi}\int_1^\infty\rho(\xi)d\xi\left(-\frac{Z\alpha}{r}\right)
e^{-2m_e\xi r}.
\end{equation}

Since hyperfine part of the Breit potential $\Delta V_B^{hfs}(r)$ is proportional 
to $\delta({\bf r})$, it is necessary to use
the reduced Coulomb Green's function with one zero argument. For this case it was obtained on
the basis of the Hostler representation after a subtraction of the pole term in
\cite{hameka}.
We represent for the sake of completeness the explicit expressions for the Green's functions, 
used in later calculations:
\begin{equation}
\label{eq:37}
\tilde G_{1S}({\bf r},0)=\frac{Z\alpha\mu^2}{4\pi}\frac{e^{-x}}{x}g_{1S}(x),~~~
g_{1S}(x)=\left[4x(\ln 2x+C)+4x^2-10x-2\right],
\end{equation}
\begin{equation}
\label{eq:38}
\tilde G_{2S}({\bf r},0)=-\frac{Z\alpha\mu^2}{4\pi}\frac{e^{-x/2}}{2x}g_{2S}(x),~~~
g_{2S}(x)=\left[4x(x-2)(\ln x+C)+x^3-13x^2+6x+4\right],
\end{equation}
where $C=0.5772...$ is the Euler constant and $x=Wr$. Then the necessary vacuum polarization corrections 
of order $\alpha^5$ to HFS of muonic ions can be presented as follows:
\begin{equation}
\label{eq:39}
\Delta E^{hfs}_{sopt~vp~1}(1S)=-E_F(1S)\frac{2\alpha}{3\pi}(1+a_\mu)\int_1^\infty
\rho(\xi)d\xi
\int_0^\infty e^{-2x\left(1+\frac{m_e\xi}{W}\right)}g_{1S}(x)dx,
\end{equation}
\begin{equation}
\label{eq:40}
\Delta E^{hfs}_{sopt~vp~1}(2S)=E_F(2S)\frac{\alpha}{3\pi}(1+a_\mu)\int_1^\infty
\rho(\xi)d\xi\int_0^\infty e^{-x\left(1+\frac{2m_e\xi}{W}\right)}g_{2S}(x)(1-\frac{x}{2})dx,
\end{equation}
where we use its designation by the index $(sopt~vp~1)$. The result of integration is in line 5 
of Tables~\ref{tb2}-\ref{tb4}.
The factor $(1+a_\mu)$ is included in \eqref{eq:39} and \eqref{eq:40}, therefore
these expressions contain corrections of orders $\alpha^5$ and $\alpha^6$. Changing $m_e\to m_1$ in
\eqref{eq:39}-\eqref{eq:40} we calculate one-loop muon vacuum polarization
contribution in second order PT of order $\alpha^6$ (see line 7 of Tables~\ref{tb2}-\ref{tb4}).

Two-loop corrections in Fig.~\ref{fig:pic2}(b,c,d,e) are of order $\alpha^6$. Let us consider
first contribution which is related with potentials \eqref{eq:29} and \eqref{eq:36},
reduced Coulomb Green's functions \eqref{eq:37}, \eqref{eq:38} and reduced Coulomb Green's
function with nonzero arguments. General structure of this contribution takes the form:
\begin{equation}
\label{eq:41}
\Delta E^{hfs}_{sopt~vp~2}=2<\psi|\Delta V^{hfs}_{1\gamma,vp}\cdot \tilde G \cdot\Delta
V^C_{vp}|\psi>.
\end{equation}
The convenient representation for reduced Coulomb Green's function with nonzero arguments was
obtained in~\cite{hameka}:
\begin{equation}
\label{eq:42}
\tilde G_{1S}(r,r')=-\frac{Z\alpha\mu^2}{\pi}e^{-(x_1+x_2)}g_{1S}(x_1,x_2),
\end{equation}
\begin{displaymath}
g_{1S}(x_1,x_2)=\frac{1}{2x_>}-\ln 2x_>-\ln 2x_<+Ei (2x_<)+\frac{7}{2}-2C-(x_1+x_2)+
\frac{1-e^{2x_<}}{2x_<},
\end{displaymath}
\begin{equation}
\label{eq:43}
\tilde G_{2S}(r,r')=-\frac{Z\alpha\mu^2}{16\pi x_1x_2}e^{-\frac{x_1+x_2}{2}}g_{2S}(x_1,x_2),
\end{equation}
\begin{displaymath}
g_{2S}(x_1,x_2)=8x_<-4x^2_<+8x_>+12x_<x_>-26x^2_<x_>+2x^3_<x_>-4x^2_>-
26x_<x^2_>+23x^2_<x^2_>-
\end{displaymath}
\begin{displaymath}
-x^3_<x^2_>+2x_<x^3_>-x^2_<x^3_>+4e^{x_<}(1-x_<)(x_>-2)x_>+4(x_<-2)x_<(x_>-2)x_>
\times
\end{displaymath}
\begin{displaymath}
\times[-2C+Ei(x_<)-\ln(x_<)-\ln(x_>)],
\end{displaymath}
where $x_1=Wr$, $x_2=Wr'$, $x_<=min(x_1,x_2)$, $x_>=max(x_1,x_2)$, $C=0.577216...$ is the Euler constant,
and $Ei(x)$ is the integral exponential function.
The substitution of \eqref{eq:29}, \eqref{eq:36}, \eqref{eq:37}, \eqref{eq:38}, 
\eqref{eq:42} and \eqref{eq:43} into
\eqref{eq:41} provides two terms for each $1S$ and $2S$ level in integral form:
\begin{equation}
\label{eq:44}
\Delta E^{hfs}_{sopt~vp~21}(1S)=-2 E_F(1S)\frac{\alpha^2}{9\pi^2}(1+a_\mu) \int_1^\infty\rho(\xi)d\xi\int_1^\infty\rho(\eta)d\eta\int_0^\infty
dx e^{-2x\left(1+\frac{m_e\eta}{W}\right)}g_{1S}(x),
\end{equation}
\begin{equation}
\label{eq:45}
\Delta E^{hfs}_{sopt~vp~22}(1S)=2 E_F(1S)\frac{\alpha^2}{9\pi^2}(1+a_\mu)\frac{16m_e^2}{W^2} \int_1^\infty\rho(\xi)\xi^2d\xi\times
\end{equation}
\begin{displaymath}
\times\int_1^\infty\rho(\eta)d\eta\int_0^\infty
x_1dx_1 e^{-2x_1\left(1+\frac{m_e\eta}{W}\right)}\int_0^\infty x_2dx_2
e^{-2x_2\left(1+\frac{m_e\xi}{W}\right)}g_{1S}(x_1,x_2),
\end{displaymath}
\begin{equation}
\label{eq:46}
\Delta E^{hfs}_{sopt~vp~21}(2S)=E_F(2S)\frac{\alpha^2}{9\pi^2}(1+a_\mu)\int_1^\infty\rho(\xi)d\xi
\int_1^\infty\rho(\eta)d\eta\int_0^\infty
\left(1-\frac{x}{2}\right)dx e^{-x\left(1+\frac{2m_e\eta}{W}\right)}g_{2S}(x),
\end{equation}
\begin{equation}
\label{eq:47}
\Delta E^{hfs}_{sopt~vp~22}(2S)=-E_F(2S)\frac{\alpha^2}{9\pi^2}(1+a_\mu)\frac{2m_e^2}{W^2} \int_1^\infty\rho(\xi)\xi^2d\xi\times
\end{equation}
\begin{displaymath}
\times\int_1^\infty\rho(\eta)d\eta\int_0^\infty
\left(1-\frac{x_1}{2}\right)dx_1 e^{-x_1\left(1+\frac{2m_e\xi}{W}\right)}\int_0^\infty
\left(1-\frac{x_2}{2}\right)dx_2
e^{-x_2\left(1+\frac{2m_e\eta}{W}\right)}g_{2S}(x_1,x_2).
\end{displaymath}
Separately, the contributions \eqref{eq:27}, \eqref{eq:28} and \eqref{eq:29},\eqref{eq:30}
are divergent but their sum is finite. Corresponding numerical values are:
\begin{equation}
\Delta E^{hfs}_{vp,vp}(1S)=
\begin{cases}
  _3^6Li:0.05~meV\\
  _3^7Li:0.20~meV\\
  _4^9Be:-0.21~meV;\\
  _5^{10}B:0.65~meV\\
  _5^{11}B:1.11~meV
 \end{cases}
 \Delta E^{hfs}_{vp,vp}(2S)=
 \begin{cases}
  _3^6Li:0.01~meV\\
  _3^7Li:0.02~meV\\
  _4^9Be:-0.02~meV.\\
  _5^{10}B:0.06~meV\\
  _5^{11}B:0.11~meV
 \end{cases}
\end{equation}
The contributions of two other amplitudes in Fig.~\ref{fig:pic2}(c,d,e) to HFS can be calculated by means 
of \eqref{eq:41}, where the replacement of the potential \eqref{eq:36}
on the following potentials should be made \cite{apm2008}:
\begin{equation}
\label{eq:48}
\Delta V^C_{VP-VP}(r)=\left(\frac{\alpha}{3\pi}\right)^2\int_1^\infty
\rho(\xi)d\xi\int_1^\infty\rho(\eta)d\eta\left(-\frac{Z\alpha}{r}\right)
\frac{1}{\xi^2-\eta^2}\left(\xi^2 e^{-2m_e\xi r}-\eta^2 e^{-2m_e\eta r}\right),
\end{equation}
\begin{equation}
\label{eq:49}
\Delta V^C_{2-loop~VP}(r)=-\frac{2Z\alpha^3}{3\pi^2 r}\int_0^1\frac{f(v)dv}{(1-v^2)}
e^{-\frac{2m_e r}{\sqrt{1-v^2}}}.
\end{equation}
Omitting further intermediate expressions we include in Tables~\ref{tb2}-\ref{tb4} total numerical values
of two-loop vacuum polarization corrections in second order PT (Fig.~\ref{fig:pic2}(b,c,d,e)) in line 9.

\begin{figure}[htbp]
\centering
\includegraphics{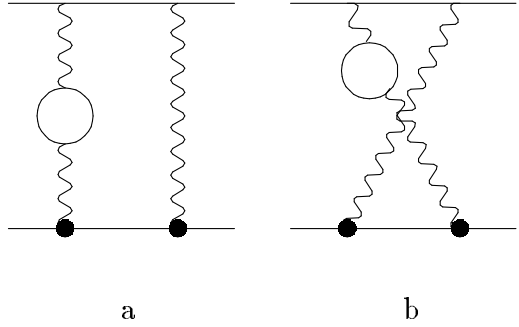}
\caption{Two photon exchange amplitudes accounting for effects of vacuum polarization and
nuclear structure. The wavy line denotes the photon. The bold point denotes the nucleus
vertex function.}
\label{fig:pic4}
\end{figure}

There is another correction for the polarization of the vacuum, which also includes the effect 
of the nuclear structure discussed in Section~II (see Fig.~\ref{fig:pic4}). To calculate it, 
it is necessary to use the potential 
$V_{2\gamma}(k)$  from \eqref{eq:25}, modifying it accordingly. As a result, the contribution to the HFS 
spectrum is determined by the following expression (the factor 2 corresponds to two exchange photons):
\begin{equation}
\label{eq:50}
E_{2\gamma,vp}^{hfs}=-\frac{2\mu^3Z^3\alpha^4}{9\pi^2n^3}
\int \frac{V_{2\gamma}(k)d^4k}{k^3}\bigl[5k^3-12m_ek^2-6(k^2-2m_e^2)\sqrt{k^2+4m_e^2}Arcth[\frac{k}{\sqrt{k^2+4m_e^2}}\bigr].
\end{equation}
Numerical integration in \eqref{eq:33} can be carried out exactly as in \eqref{eq:25}
(line 12 of Tables~\ref{tb2}-\ref{tb4}). The contribution of muon VP in $2\gamma$-amplitudes
with the nuclear structure is written in line 13 of Tables~\ref{tb2}-\ref{tb4}.

In order to increase the accuracy of the calculation we consider also hadron vacuum polarization 
(HVP) contribution which arises, like the electron polarization of the vacuum, in the first order PT, 
in the second order PT, and in two-photon exchange amplitudes. To obtain it we use 
a standard replacement in photon propagator of the form \cite{faustov1999}.
\begin{equation}
\label{eq:51}
\frac{1}{k^2}\to\left(\frac{\alpha}{\pi}\right)\int_{s_{th}}^\infty\frac{\rho_{had}(s)ds}{k^2+s},
~~~\rho_{had}(s)=\frac{(s-4m_\pi^2)^{3/2}}{12s^{5/2}}|F_\pi(s)|^2,
\end{equation}
where $F_\pi(s)$ is the pion form factor. Total hadron vacuum polarization contribution is
presented in Tables~\ref{tb2}-\ref{tb4} (line 14).

\section{Radiative corrections to two photon exchange diagrams.}

\begin{figure}[htbp]
\centering
\includegraphics{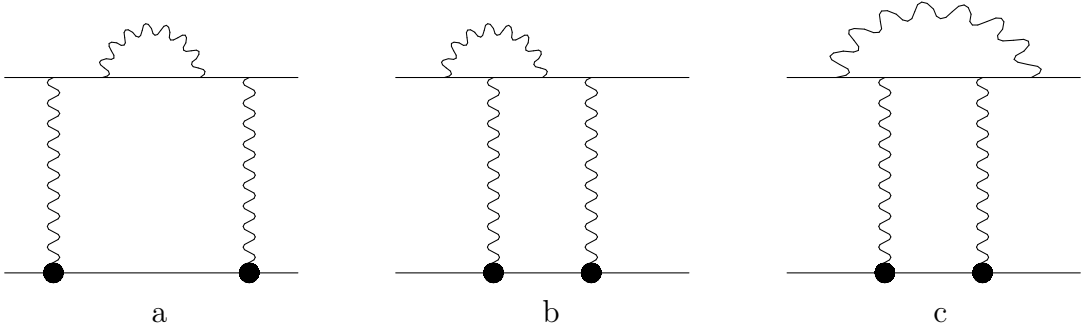}
\caption{Direct two-photon exchange amplitudes with radiative corrections to muon line
giving contributions of order $E_F\alpha(Z\alpha)$ to the hyperfine structure. Wave line on the diagram denotes
the photon. Bold point on the diagram denotes the vertex operator of the nucleus.}
\label{fig:pic6}
\end{figure}

The results already obtained in the Tables~\ref{tb2}-\ref{tb4} clearly show that the corrections 
to the structure of the nucleus are dominant. In this connection, it seems useful 
to consider another correction for the structure of the nucleus of order $\alpha^6$
shown in Fig.~\ref{fig:pic6} to refine the results. The amplitudes of two-photon exchange with 
radiative corrections to the muon line can be calculated in the framework of the 
calculation method formulated in Section~II. For a radiative photon, the 
Fried-Yennie gauge is used, in which each of the amplitudes in Fig.~\ref{fig:pic6}
(muon self-energy, muon vertex correction, amplitude with the spanning photon) 
can be represented by a finite integral expression. The general structure of the
amplitudes in Fig.~\ref{fig:pic6} is the following:
\begin{equation}
\label{eq:52}
{i\cal M}=\frac{(Z\alpha)^2}{\pi^2}\int d^4k
\left[
\bar u(q_1)L_{\mu\nu}u(p_1)\right]D_{\mu\omega}(k)D_{\nu\lambda}(k)
\left[\bar v_\rho(p_2){\cal O}_{\rho\omega\beta}
{\cal D}_{\beta\tau}(p_2+k){\cal O}_{\tau\lambda\alpha}v_\alpha(q_2)\right],
\end{equation}
where the vertex operator ${\cal O}_{\rho\omega\beta}$ describing the
photon-nucleus interaction is determined by the nucleus electromagnetic form factors as in \eqref{eq:8}
for the nucleus of spin $3/2$.

The spin $3/2$ particle propagator and the photon propagator in the Coulomb gauge are equal to
\begin{equation}
\label{eq:53}
{\cal D}_{\alpha\beta}(p)=\frac{\hat p+m_2}{p^2-m_2^2+i0}\left[g_{\rho\beta}-\frac{1}{3}\gamma_\rho\gamma_\beta
-\frac{2p_\rho p_\beta}{3m_2^2}-\frac{\gamma_\rho p_\beta-\gamma_\beta p_\rho}{3m_2}\right],
\end{equation}
\begin{equation}
\label{eq:54}
D_{\lambda\sigma}(k)=\frac{1}{k^2+i0}\left[g_{\lambda\sigma}+\frac{k_\lambda k_\sigma-k_0 k_\lambda g_{\sigma 0}-
k_0 k_\sigma g_{\lambda 0}}{{\bf k}^2}\right].
\end{equation}
The lepton tensor $L_{\mu\nu}$ is equal to a sum of three terms coming from three amplitudes in 
Fig.~\ref{fig:pic6}: 
\begin{equation}
\label{eq:55}
L_{\mu\nu}=L_{\mu\nu}^{se}+L_{\mu\nu}^{vertex}+L_{\mu\nu}^{jellyfish}.
\end{equation}
All three terms of the lepton tensor were obtained in integral form and are written explicitly 
in \cite{apm2,eides1,eides2,eides3}.

The construction of hyperfine potential by means of amplitudes in Fig.~\ref{fig:pic6} in the case
of the spin 3/2 nucleus can be performed by the method of projection operators as in section~II.
Neglecting the recoil effects in the denominator of the nucleus propagator we obtain
that a sum of direct and crossed amplitudes is proportional to $\delta(k_0)$:
\begin{equation}
\label{eq:56}
\frac{1}{2m_2k_0+i0}+\frac{1}{-2m_2k_0+i0}=-\frac{i\pi}{m_2}\delta(k_0).
\end{equation}

As a result three types of contributions of order $E_F\alpha(Z\alpha)$ to HFS of muonic ions
of lithium, beryllium and boron are expressed in integral form over the loop momentum $k$ and the Feynman parameters:
\begin{equation}
\label{eq:57}
\Delta E^{hfs}_{se}=E_F6\frac{\alpha(Z\alpha)}{\pi^2}\int_0^1 xdx\int_0^\infty
\frac{G_E(k^2)G_M(k^2)dk}{x+(1-x)k^2},
\end{equation}
\begin{equation}
\label{eq:58}
\Delta E^{hfs}_{vertex-1}=-E_F24\frac{\alpha(Z\alpha)}{\pi^2}\int_0^1 dz\int_0^1 xdx\int_0^\infty
\frac{G_E(k^2)G_M(k^2)\ln [\frac{x+k^2z(1-xz)}{x}]dk}{k^2},
\end{equation}
\begin{equation}
\label{eq:59}
\Delta E^{hfs}_{vertex-2}=E_F8\frac{\alpha(Z\alpha)}{\pi^2}\int_0^1 dz\int_0^1 dx\int_0^\infty\frac{dk}{k^2}
\biggl\{\frac{G_E(k^2)G_M(k^2)}{[x+k^2z(1-xz)]^2}\bigl[-2xz^2(1-xz)^2k^4+
\end{equation}
\begin{displaymath}
zk^2(3x^3z-x^2(9z+1)+x(4z+7)-4)+x^2(5-x)\bigr]-\frac{1}{2}\biggr\},
\end{displaymath}
\begin{equation}
\label{eq:60}
\Delta E^{hfs}_{jellyfish}=E_F4\frac{\alpha(Z\alpha)}{\pi^2}\int_0^1(1-z)dz\int_0^1(1-x)dx\int_0^\infty
\frac{G_E(k^2)G_M(k^2)dk}{[x+(1-xz)k^2]^3}
\end{equation}
\begin{displaymath}
\times\bigl[6x+6x^2-6x^2z+2x^3-12x^3z-12x^4z+k^2(-6z+18xz+4xz^2+7x^2z-30x^2z^2-
\end{displaymath}
\begin{displaymath}
2x^2z^3-36x^3z^2+12x^3z^3+24x^4z^3)+k^4(9xz^2-31x^2z^3+34x^3z^4-12x^4z^5\bigr]
\end{displaymath}
All contributions \eqref{eq:57}-\eqref{eq:60} are expressed in terms of electric and dipole magnetic
form factors.
The term $1/2$ in figure brackets~\eqref{eq:59} is related to the subtraction term of the
quasipotential. All corrections \eqref{eq:57}, \eqref{eq:58}, \eqref{eq:59}, \eqref{eq:60}
are expressed through the convergent integrals. Numerical results for the corrections
\eqref{eq:57}-\eqref{eq:60} are presented in Tables~\ref{tb2}-\ref{tb4} (line 17).

\section{Conclusion}

In this work we carry out a calculation of S-states hyperfine splittings in a number of muonic ions.
We consider that light muonic ions of the lithium, beryllium and boron can be
used in experiments of the CREMA collaboration. 
Our precise calculation of the HFS includes taking into account the various corrections of the 
fifth and sixth orders in $\alpha$, which were previously taken into account also in the 
study of the hyperfine structure of the spectrum of other muonic atoms \cite{apm1,apm2,apm3,apm4}. 
One significant 
difference between these calculations and the previous ones is due to the fact that 
in this paper we investigate the nuclei of spin 1, 3/2, 3. 
For spin 3/2 nuclei we have included the effects of two-photon interactions 
in the framework of quantum electrodynamics of spin particles 3/2. Corrections 
to the structure of the nucleus, which 
are determined by two-photon exchange amplitudes, as follows from the results 
of the Tables~\ref{tb2}-\ref{tb4}, play a very important role in achieving high accuracy of calculation. 
They are defined in our
approach by the electromagnetic form factors of the nuclei, which in this case must be taken 
from experimental data. Intensive experimental studies of the scattering of leptons 
by light nuclei were carried out several dozen years ago. The results obtained then are reflected in \cite{scat,fuller}. 
We use these results, although the accuracy of determining all the required form factors 
is not very high, as would be desirable. 
For this reason, we use different parameterizations (Gaussian, uniformly charged sphere) 
for form factors and compared the numerical results for them to understand how they can differ. 
Complete numerical values for hyperfine splittings of S-levels are presented 
in the Tables~\ref{tb2}-\ref{tb4} for the Gaussian parameterization.
Numerous corrections for vacuum polarization are taken into account in the traditional way, 
which is connected with the modification of the photon propagator \cite{t4}.

\begin{figure}[htbp]
\centering
\includegraphics{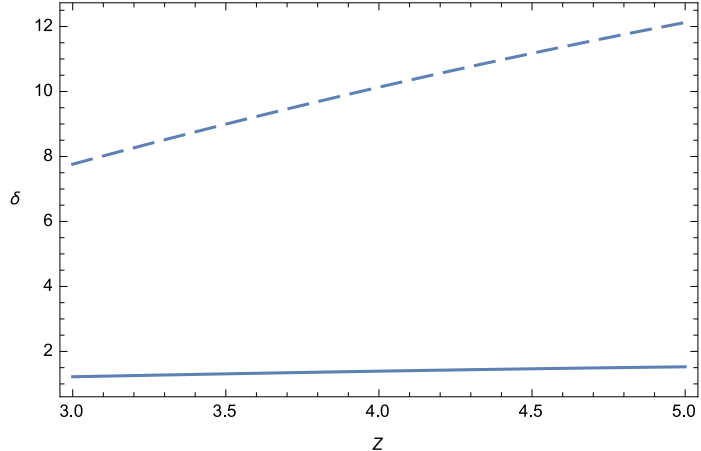}
\caption{Relative order contributions $\delta$ in percent of vacuum polarization (solid line, order $\alpha^5$) and nuclear structure (dashed line, order $\alpha^5$) to hyperfine
structure of muonic ions of lithium, beryllium and boron.}
\label{fig:delta}
\end{figure}

We present in Tables~\ref{tb2}-\ref{tb4} all obtained results for a calculation of corrections
in first and second orders of perturbation
theory. In the text of our work we give numerous references on that results indicating the lines of
Tables~\ref{tb2}-\ref{tb4} where these corrections are presented.
The dependence of basic corrections of order $\alpha^5$ on the nucleus is shown in Fig.~\ref{fig:delta}.
As pointed out above the hyperfine structure of muonic ions of lithium, beryllium and boron 
was investigated previously in \cite{drake}. The authors of \cite{drake} gave only estimates of basic
contributions in hyperfine structure. In this work we make an attempt to improve their results
accounting for different corrections.

The results of calculating various corrections are presented with an accuracy of 0.01 meV. 
A number of corrections for the polarization of a vacuum have precisely this order. 
But this does not mean that the accuracy of our calculation is so high. 
There are already mentioned above corrections to the structure of the nucleus which give the main theoretical uncertainty in the total obtained results. This uncertainty, due to the electromagnetic form factors 
of the nuclei, can be about 1 percent of the correction to the structure of the nucleus of the order 
$\alpha^5$. Thus, we estimate approximately the errors in the calculation of the HFS spectrum in the form:
$\delta E^{hfs}(^6_3Li)=\pm 1$ meV, $\delta E^{hfs}(^7_3Li)=\pm 4$ meV, 
$\delta E^{hfs}(^9_4Be)=\pm 4.5$ meV, $\delta E^{hfs}(^{10}_5B)=\pm 14$ meV,
$\delta E^{hfs}(^{11}_5B)=\pm 24$ meV.
It is appropriate to note here that corrections to the structure of the nucleus of order $\alpha^5$ (the Zemach correction) significantly exceed all other corrections listed in the Tables~\ref{tb2}-\ref{tb4}
(see Fig.~\ref{fig:delta}). 
We can say that in this respect the hyperfine splitting differs from the Lamb shift $(2P_{1/2}-2S_{1/2})$
in which the correction of the leading order to the structure of the nucleus and the correction 
for one-loop vacuum polarization are comparable in magnitude and have different signs. 
Thus, a precise measurement of the HFS in muonic ions of lithium, beryllium and boron, taking 
into account the obtained theoretical results, will allow obtaining more accurate values of the Zemach 
radius for these atoms.

There is another correction for the polarizability of the nucleus, which
is not considered in this paper. In the case of muonic deuterium, this correction was calculated in \cite{ibk1,ibk2}. 
For tritium and helium-3 nuclei, this type of correction was investigated in \cite{friar1}. 
The correction for the polarizability of the nucleus is determined by the interaction 
of a multinucleon system with an external electromagnetic field, as a result of which 
the nucleus passes into an excited state. In \cite{kp2007}, 
general expressions were obtained for calculating the various parts of the correction for the 
nuclear polarizability in the HFS spectrum. 
Another approach to solving this problem is 
connected with the use of the dispersion method, in which the correction for the nuclear 
polarizability is determined by known general formulas and is expressed in terms of the 
spin-dependent structure functions of the nucleus. If such spin-dependent structure 
functions of the nuclei were measured exactly experimentally as electromagnetic form 
factors, then they could be used in calculations. Otherwise, we must consider the motion 
of the nucleons of the nucleus in the effective potential field and their interaction 
with an external field. In the case of the Lamb shift, such calculations 
were performed in \cite{bacca}.
In Refs \cite{ibk1,friar1,kp2007}, the correction for the polarizability in the HFS 
was discussed together with effects on the structure of the nucleus, so that the total 
correction was represented in the form: $\delta E^{hfs}=\delta E^{hfs}_{Low}+\delta E^{hfs}_{Zemach}+
\delta E^{hfs}_{pol}$. In our approach, in which we use the electromagnetic form factors of the nucleus, 
the correction for the nuclear structure
of order $\alpha^5$ (lines 4-5) corresponds to the sum $\delta E^{hfs}_{Low}+\delta E^{hfs}_{Zemach}$, 
in calculating which the nucleus is represented as the sum of nucleons. The correction 
for the polarizability is of order $O(Z\alpha m_1/m_2)$, so its possible numerical value for different nuclei 
(0.6 meV $(^6_3Li$), 1.8 meV $(^7_3Li$), -1.6 meV $(^9_4Be$), 4.7 meV $(^{10}_5B$), 7.3 meV $(^{11}_5B$))
is comparable in magnitude to those errors that are connected with errors in measuring nuclear 
form factors. At the same time, it should be noted that the correction for the polarizability for a deuteron substantially exceeds this estimate. Therefore, its exact calculation becomes a very urgent problem. 
Our work in this direction is in progress.

\begin{acknowledgments}
The work is supported by Russian Science Foundation (grant No. RSF 18-12-00128) and
Russian Foundation for Basic Research (grant No. 18-32-00023) (F.A.M.).
\end{acknowledgments}

\end{document}